\documentclass[%
 reprint,
 superscriptaddress,
 amsmath,
 amssymb,
 aps,
]{revtex4-2}

\usepackage{pdfpages}
\makeatletter
\AtBeginDocument{\let\LS@rot\@undefined}
\makeatother
\usepackage{graphicx}
\usepackage{dcolumn}
\usepackage{bm}
\usepackage{overpic}
\usepackage{hyperref}

\usepackage{color}

\usepackage{mathrsfs}

\begin{document}

\preprint{APS/123-QED}

\title{Self-organized quantization and oscillations on continuous fixed-energy sandpiles}

\author{Jakob Niehues}
  \affiliation{Niels Bohr Institute, Copenhagen University, Blegdamsvej 17, 2100 Copenhagen, Denmark}
\author{Gorm Gruner Jensen}%
  \email{gruner.jensen@nbi.ku.dk}
  \affiliation{Niels Bohr Institute, Copenhagen University, Blegdamsvej 17, 2100 Copenhagen, Denmark}
\author{Jan O. Haerter}%
  \email{haerter@nbi.ku.dk}
  \affiliation{Niels Bohr Institute, Copenhagen University, Blegdamsvej 17, 2100 Copenhagen, Denmark}
  \affiliation{Leibniz Centre for Tropical Marine Research, Fahrenheitstrasse 6, 28359 Bremen, Germany}
  \affiliation{Jacobs University Bremen, Campus Ring 1, 28759 Bremen, Germany}

\date{\today}

\begin{abstract}
\noindent
Atmospheric self-organization and activator-inhibitor dynamics in biology provide examples of checkerboard-like spatio-temporal organization.
We study a simple model for local activation-inhibition processes.
Our model, first introduced in the context of atmospheric moisture dynamics, is a continuous-energy and non-Abelian version of the fixed-energy sandpile model.
Each lattice site is populated by a non-negative real number, its energy.
Upon each timestep all sites with energy exceeding a unit threshold re-distribute their energy at equal parts to their nearest neighbors.
The limit cycle dynamics gives rise to a complex phase diagram in dependence on the mean energy $\mu$:
For low $\mu$, all dynamics ceases after few re-distribution events.
For large $\mu$, the dynamics is well-described as a diffusion process, where the order parameter, spatial variance $\sigma$, is removed.
States at intermediate $\mu$ are dominated by checkerboard-like period-two phases which are however interspersed by much more complex phases of far longer periods.
Phases are separated by discontinuous jumps in $\sigma$ or $\partial_{\mu}\sigma$ --- akin to first and higher-order phase transitions.
Overall, the energy landscape is dominated by few energy levels which occur as sharp spikes in the single-site density of states and are robust to noise.
\end{abstract}

\keywords{Suggested keywords}
\maketitle


\section{\label{sec:intro}Introduction\protect}
\noindent
The deterministic fixed-energy sandpile (DFES) model  \cite{tang1988critical}, also known as chip-firing game \cite{bjorner1991chip}, is a variant of the classical Bak-Tang-Wiesenfeld (BTW) sandpile model \cite{BTW}, where the driving and dissipation are replaced by constant energy and closed boundary conditions.
It consists of a non-negative integer-valued field on a network that is updated in discrete time-steps according to a deterministic toppling rule.
Studies have found an absorbing-state phase transition, with strong dependence on the initial conditions, between active and inactive states \cite{Vespignani_1998, Vespignani_2000, Munoz_2001, Dickman_1998, Dickman_2000, Fey_2010}, Abelian dynamics of the absorbing phase \cite{Dhar90, BAL92, FdBMR09, FdBMR09} and short-period attractors in the active phase \cite{Bagnoli_2003, Dall_Asta_2006}.
For semantic consistency, in the following we maintain the term "energy" to refer to the abstract quantity located on each site.
A variant of the original BTW sandpile model that replaces the integer-valued field by a continuous one, while keeping a dissipative boundary and the slow insertion of energy into the system, has been suggested by Yi-Cheng Zhang \cite{Zhang}. In addition to the celebrated self-organized criticality of its discrete counterpart it also shows a self-organized quantization effect that leads to a peaked energy landscape.

We explore an energy conserving model \cite{Di-urnal} on networks or lattices, where a site's entire energy, a non-negative real number, is distributed to its neighbors at equal shares when the site's energy exceeds a numerical threshold.
All sites are updated synchronously.
Since the amount of energy that is distributed from a site is all of that site's energy, the dynamics is non-Abelian.

We simulate the dynamics long enough for the system to reach stationary limit cycles, which are usually periodic orbits. We investigate the emerging spatio-temporal structure which we characterize by few order parameters and construct phase diagrams to show how the order parameters depend on the average energy $\mu$ and the network geometry.
We observe self-organized numerical discretization of the energy field, strong non-ergodicity, periodic limit cycle oscillations, stochastic phase transitions, and complex spatio-temporal topologies.
These findings partially resemble the behavior of discrete DFES, but show a richer spectrum of behavior.


When initially introduced \cite{Di-urnal}, the model was physically motivated: 
\citeauthor{chen1997diurnal} (\citeyear{chen1997diurnal}) observed a bi-diurnal oscillation of cloudiness over the Western Pacific Warm Pool, a phenomenon they referred to as "diurnal dancing." 
They attributed this checkerboard-like spatio-temporal pattern to the formation of large thunderstorm cloud patches, termed "mesoscale convective systems." 
These mesoscale convective systems are each day set off by the solar shortwave diurnal radiation cycle.
Yet, as the mesoscale convective systems on one day dissipate, they suppress their further activity by dry and often cold downbursts.
Such dense downbursts, often referred to as ``cold pools,'' form in the atmospheric boundary layer when rain evaporates beneath thunderstorm clouds \cite{Characterizing}. 
When cold pools reach the surface, a horizontal flow is forced \cite{moisture, vaporrings, StructureOutflows, GravityCurrent} and acts to re-distribute moisture laterally to neighboring regions. 

From simulations it is known that such cold pool induced moisture re-distribution can trigger new convective events \cite{Haerter_2019,Di-urnal}.
Cold pools therefore play a key role in the structural organization of rain cells \cite{nissen2021weakened, Di-urnal, ConSelfAgg, CSA_CP_domainsize, objectbased,RoleofCP}.
Under certain conditions, where there is a temporally periodic, daily-repeating, surface temperature forcing, extended cloud and precipitation patches form in sub-regions of the simulated domain.
These patches show a dynamics where moisture is pushed out laterally by cold pools, causing the sub-cloud environment to become dry, i.e., inactive, and its surroundings to become humid, thus potentially active.
The periodic forcing acts to set a "clock" for the dynamics which, in regions that are sufficiently humid, unfolds almost simultaneously throughout the simulated region.
This dynamics is the essence of the model we explore here,
where we simplify by representing each potential thunderstorm location by a lattice site. 
If the moisture on a site exceeds a certain threshold, it triggers a "toppling" event, which causes all moisture to be distributed to the neighboring sites during a discrete timestep.

Checkerboard-like self-organized patterns are also common in biology: 
the Notch-Delta activation-inhibition mechanism \cite{artavanis1999notch} has been modeled as a lattice gas cellular automaton \cite{deutsch2005cellular}.
In this model, each site on a 2D lattice represents a tissue cell, which has a concentration of Notch and Delta, and cells communicate with their nearest neighbors: 
Notch interacts with Delta by activating Delta within neighboring cells but inhibits Delta locally. 
In turn, Notch can be activated by high delta concentration in neighboring cells.
Under specific interaction strength a checkerboard pattern of cells with high and low levels of Notch and Delta can result. 
Under additional noise, states deviating from a perfect checkerboard pattern can emerge, referred to as having "frustrated furrows." 
In these "frustrated furrows" regions of intermediate concentrations of Notch are present, which we call furrows.
Spatio-temporal striped or checkerboard-like patterns were previously obtained in probabilistic reaction-diffusion models, where reactions took place stochastically \cite{deutsch2005cellular}.
Patterns were further substantially influenced by the lattice geometry.

The system we discuss carries randomness only in the initial condition, that is, the initial energy value on each site. 
However, we explicitly focus on effects that {\it do not} depend on the details of this initial condition, demanding only that the probability distribution of initial site energies be continuous.
The dynamics is otherwise deterministic and evolves to a set of attractors with typical characteristics which depend on the average energy $\mu$ and are robust to noise.


\section{\label{sec:model}Model and Methods}
\noindent
To each site $i$ of a lattice of $N$ sites, such as a 2D square, triangular, or honeycomb lattice, we assign a non-negative real number $z_i\in\mathbb{R}_0^+$, representing the energy of the site.
The time-dependent $N$-dimensional \textit{system state} $Z(t) = \left\{z_i(t)\right\}$ is updated in integer timesteps $t\in\{0,1,2,\ldots\}$, which can be stated as an operator equation
\begin{equation}
\label{eq:updating_operator}
     Z(t+1) = \hat{T} Z(t),
\end{equation}
where $\hat{T}$ is an operator that updates the system according to the toppling rule:
all sites $i$ with $z_i$ exceeding a unit threshold ("active sites") simultaneously distribute all their energy at equal shares to their nearest neighbors, that is,

\begin{equation}
    \label{eq:toppling_rule}
    z_i(t+1) = z_i(t) \cdot \left[1 - \Theta \left(z_i(t) - 1 \right) \right]
    + \sum_{j\in\mathcal{N}_i} \frac{z_j(t)}{k_j} \Theta \left(z_j(t) - 1 \right),
\end{equation}
where $\mathcal{N}_i$ is the set of nearest neighbors of $i$, $k_j$ refers to the number of neighbors of $j$, and $\Theta$ is the Heaviside step function, defined as
\begin{equation*}
    \Theta(x) \equiv 
    \Bigg\{
    \begin{array}{cc}
        0 & \text{if } x \le 0 \\
        1 & \text{if } x > 0\,,
    \end{array}
\end{equation*}
where $x\in\mathbb{R}$.
Note that, as with the discrete DFES, the lack of driving and dissipation ensures the dynamical conservation of the average energy per site, $\mu$,
\begin{equation*}
    \mu \equiv \frac{1}{N} \sum_{i=1}^N z_i\,,
\end{equation*}
a parameter that characterizes a realization of the model on a certain lattice.

We initialize the model with random initial conditions, that is, each site energy is sampled from a continuous probability density function, independent of the energy values of all other sites.
We compute timeseries of states as a discrete map which we analyze in terms of limit cycles and transient behavior (Sec.~\ref{sec:results}).
We focus on the 2D square lattice, but test the robustness of our findings against boundary-induced frustration (Fig.~S1), geometry (Figs. S3 and S4), dimensionality (Figs. \ref{fig:3_lin_pd}, \ref{fig:4_lin_states} and S4) and noise (Fig.~S7).
Figures with prefix S are located in the supplement.

\subsection{Basic model features}
\noindent
A key property of the discrete DFES is that its dynamics is Abelian in the absorbing phase:
If the activity comes to a halt after a finite time, both the energy of each site in the final (absorbing) state and how often each site topples during the transient period are independent of the order in which single sites are updated according to
\begin{equation}
    \label{eq:DFES_toppling_rule}
    n_i(t+1) = n_i(t) - k_i \cdot \Theta  \left(n_i(t) - k_i \right)
    + \sum_{j\in\mathcal{N}_i} \Theta \left(n_j(t) - k_j \right),
\end{equation}
i.e. a toppling site gives one discrete grain to each neighbor \cite{Dhar90, BAL92, FdBMR09, FdBMR09}.
The differences between Eqs~\ref{eq:toppling_rule} and \ref{eq:DFES_toppling_rule} are three-fold: 
(i) the non-negative real-valued energy field $z_i(t)$ is replaced by a non-negative integer energy field $n_i(t)$; (ii) the threshold is rescaled from 1 to $k_i$; and (iii) the amount of energy that is re-distributed upon toppling is now the constant $k_i$ instead of the toppling site's energy itself.
Hence, for the DFES, a site does not have to be empty after toppling, even if the neighbors do not topple.
If we were to transfer the amount of energy $1/k_i$ to each neighbor, our model would be a rescaled version of the Abelian DFES with the addition of constant noise that has no effect on the dynamics. See \cite{FdBMR09, Dhar98} for a proof of the Abelian property of DFES.

If single sites are updated successively, our continuous sandpile model is non-Abelian, because the amount of energy distributed upon toppling of a site equals the site's energy, whereas in the discrete model, the neighbors of a toppling site receive a fixed energy quantity.
To exemplify the non-Abelian property, consider a $3\times3$ or larger square lattice with periodic boundaries.
Let all sites carry zero energy except for two neighboring sites that are equally filled above the threshold, say with an energy of two.
The system will relax into an inactive state after one update, the exact form of which depends on the order of toppling.
Since our model is non-Abelian, we have to be specific about the update rule to avoid ambiguity.
We choose to study the model with simultaneous updates to ensure that the rule is both symmetric and deterministic.
It is symmetric in the sense that no sites or directions are given any special priority, and deterministic in the sense that any given system state has a unique successor according to Eq.~\ref{eq:toppling_rule}.
Note that the mapping is not unique when time is reversed:
timeseries, which are sequences of states generated by updating the system, are highly asymmetric in time, because a given system state in general does not have a unique predecessor.
Due to its Abelian property the discrete DFES also does not have time-reversal symmetry in the absorbing phase.

\begin{figure}
    \centering
    \includegraphics{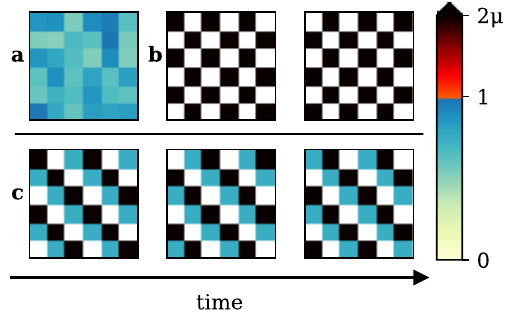}
    \caption{Possible periodic attractors with $\mu = 0.75$ on a $6\times 6$ square lattice:
    (a) inactive state with period 1;
    (b) period-two "checkerboard";
    (c) period-three "diagonal wave".}
    \label{fig:1_SimpleExample}
\end{figure}

It is in principle possible that a timeseries is periodic with period $T \in \mathbb{N}$ in the sense that $Z(t+T) = Z(t)\quad \forall t$.
Here we are interested in the smallest positive integer number $T$ with this property.
When initializing the model with random initial conditions the resulting timeseries may not be strictly periodic (up to finiteness of its implementation on a computer). 
The dynamics may however converge towards a periodic orbit (a sequence of states), which constitutes an attractor in phase space and which we refer to as a {\it limit cycle}.
The model may give rise to multiple distinct limit cycle oscillations for a given mean energy $\mu$,
e.g., the system in Fig.~\ref{fig:1_SimpleExample} shows possible periodic attractors with $\mu = 0.75$.
Fig.~\ref{fig:1_SimpleExample}a shows an inactive absorbing state, that is, a state can never be left once it is entered.
Note that the only active absorbing state is a homogeneous state in which all sites have the same threshold-exceeding energy $\mu$.
The simplest period-two attractor is the spatially and temporally anti-correlated "checkerboard" on a bipartite lattice (Fig.~\ref{fig:1_SimpleExample}b).
It consists of two sublattices with energies 0 and $2\mu$ on each site respectively that alternate between one timestep and the next.
If $L$ is even, boundary conditions are commensurate with period two and a checkerboard is possible.
If $L$ is odd, bipartition is not possible and a departure from a perfect checkerboard must result.
Fig.~\ref{fig:1_SimpleExample}c shows a simple example of a period-three limit cycle.

The characteristics of the periodic attractors in the limits of high and low $\mu$ can be inferred by heuristic arguments:
For $\mu \ll 1$, all activity is expected to come to a halt after a transient:  all energy is diffused in such a way that no active sites are left and the system reaches an inactive absorbing state.
For $\mu \gg 1$, the limit cycle is expected to be maximally active and homogeneous, because almost all sites topple during the first update and the operator $\hat{T}$ in Eq.~\ref{eq:updating_operator} effectively acts as a diffusion operator. 
For intermediate $\mu$ the combination of incessant activity and non-homogeneous pattern formation is possible.

In terms of $\mu$ at least one phase transition, an absorbing-state phase transition, is expected between the absorbing inactive ($\mu\ll 1$) and active ($\mu\approx 1$) phase.
Note that there is at least one site that never topples, if an inactive absorbing attractor state is reached.
This can be proved by contradiction (following the literature for the BTW DFES \cite{FdBMR09, FLP10}):
Suppose the activity stops after every site has toppled at least once, then the first site to stop toppling, say at time $t=t_s$ receives enough energy by its toppling neighbors at times $t \ge t_s$ to topple again, a contradiction.
Assuming the same connectivity $k$ for all sites of a finite lattice, this condition becomes an equivalence: 
The activity must come to a halt if there is at least one site that never topples, because all of its $k$ neighbors can maximally topple a total amount of $k$ times altogether. Repeating this argument leads to a finite total number of toppling events for the whole lattice which must result in an inactive absorbing state.

\subsection{Order parameters}
\noindent
The activity density $a(t)$ at time $t$ is defined as the fraction of active sites
\begin{equation}
    a(t) \equiv \frac{1}{N} \sum_i \Theta\left( z_i(t) - 1\right).
\end{equation}
$a(t)$ equals zero for inactive system states. A "checkerboard" would have $a(t)=1/2$ and $a(t)=1$ would require all sites to be active.\\
The normalized (spatial) standard deviation at time $t$ is given by
\begin{equation}
    \sigma(t) \equiv \frac{1}{\sqrt{N} \mu}\sqrt{\sum_{i} \left( z_i(t) - \mu \right)^2 }\,.
\end{equation}
$\sigma(t)$ equals zero for homogeneous system states and unity for "checkerboards."
A random initial system state drawn from a  mean-$\mu$ uniform distribution from 0 to $2\mu$ [exponential distribution] has expectation values $a(0) = 1 - 1/(2\mu)$ $\left[\exp(-1/\mu)\right]$ and $\sigma(0) = 1/\sqrt{3} \approx 0.577$ $\left[\mu\right]$.

A timeseries has converged to a period-$T$ limit cycle, if a transient time $t^*$ and a period $T$ with $Z(t) = Z(t+T)\quad \forall t \ge t^*$ exist.
For a period-$T$ limit cycle, the mean activity density $\langle a \rangle\equiv \sum_t^{t+T-1}a(t)/T$ and mean normalized standard deviation $\langle \sigma \rangle\equiv \sum_t^{t+T-1}\sigma(t)/T$ are computed by averaging over one complete cycle of length $T$, i.e. a sequence of $T$ states.

\subsection{Parameter scan}\label{sec:numerical_simulations}
\noindent
Simulations were carried out to numerically identify limit cycles for realizations of the model on a given lattice. 
An overview of simulation parameters can be found in Tab.~\ref{tab:simulation_parameters}.
\begin{table*}[]
    \centering
    \begin{ruledtabular}
    \begin{tabular}{llllllllll}
        Fig.&geometry&$k$&$N$&distr.&$\mu_{\text{min}}$, $\mu_{\text{max}}$, $\delta \mu$&$\log_{10} t_{\text{sim}}$&$\log_{10}\left(T_{\text{max}}/5 \right)$&$s$&$s^*_\text{min}$, $s^*_\text{max}$, $f_{s^*}$ \\\hline
        \ref{fig:3_lin_pd}, \ref{fig:4_lin_states}&linear&2&1997&uni.&0.5, 10.5, 0.02&3, 4, 5, 6&1, 2, 3, 3&$1.07\cdot 10^{-13}$&$3\cdot10^{-1}$, $5\cdot10^{-1}$, 4/3\\\hline
        \ref{fig:2_transient}&square&4&$10\times 10$&uni.&$\mu=2.1$&&&\\\hline
        \ref{fig:5_squ_pd}&square&4&$31\times 31$&uni.&0.5, 5.0, 0.01&3, 4, 5, 6, 7, 8&1, 2, 3, 4, 5, 5&$8.54\cdot10^{-13}$&$ 8.53\cdot10^{-13}$, $10^{-2}$, $10^3$\\\hline
        S1&square&4&$100\times 100$&uni.&0.5, 5.0, 0.01&3, 4, 5, 6&1, 2, 3, 3&$8.88\cdot10^{-12}$&$10^{-14}$, $10^{-3}$, $10^4$\\\hline
        \ref{fig:5_squ_pd}, S1&square&4&$101\times 101$&uni.&0.5, 5.0, 0.01&3, 4, 5, 6, 7&1, 2, 3, 3, 3&$9.06\cdot 10^{-12}$&$9\cdot10^{-12}$ , $1\cdot10^{-1}$, $10^3$\\\hline
        S2&square&4&$101\times 101$&exp.&0.5, 5.0, 0.01&3, 4, 5, 6, 7&1, 2, 3, 3, 3&$9.06\cdot 10^{-12}$&$9\cdot10^{-12}$, $10^{-3}$, $10^3$\\\hline
        \ref{fig:7_squ_101_noise}, S7&square&4&$101\times101$&uni.&0.5, 5.0, 0.01&$t_\text{sim}\!\!=$233,100&&&\\\hline
        S7 top&square&4&$101\times101$&uni.&0.5, 5.0, 0.01&$t_\text{sim}\!\!=$233,100&2&$3.00\cdot10^{-2}$&$10^{-2}$, $3\cdot10^{-1}$, 10/3\\\hline
        S6&square&4&$101\times 101$&uni.&0.625, 0.775, 0.0005&3, 4, 5&1, 2, 3&$9.06\cdot 10^{-12}$&$10^{-12}$, $10^{-11}$, $10^3$\\\hline
        \ref{fig:5_squ_pd}, \ref{fig:6_squ_states}&square&4&$307\times 307$&uni.&0.5, 5.0, 0.01&3, 4, 5&3, 3, 3&$8.37\cdot 10^{-11}$&$7\cdot10^{-2}$, $1\cdot10^{-1}$, 8/7\\\hline
        S3&honeycomb&3&$100\times100$&uni.&0.5, 5.0, 0.01&3, 4, 5, 6, 7&1, 2, 3, 3, 3&$8.88\cdot 10^{-12}$&$10^{-12}$, $10^{-3}$, $10^3$\\\hline
        S4&linear&4&1997&uni.&0.6, 1.8, 0.0025&3 ,4, 5, 6&1, 2, 3, 3&$1.77\cdot 10^{-12}$&$10^{-8}$, $4\cdot10^{-1}$, $10^3$\\\hline
        S4, S5&triangular&6&$101\times101$&uni.&0.5, 1.7, 0.005&3, 4, 5, 6, 7&1, 2, 3, 3, 3&$9.06\cdot 10^{-12}$&$10^{-14}$, $10^{-3}$, $10^3$
        
    \end{tabular}
    \caption{Parameters used in the different simulations.
    If several values for $t_\text{sim}$ and $T_\text{max}$ are listed, step \ref{itm:increments} was repeated that many times.
    To give an example, in the simulation on the $k=2$ linear ring lattice step \ref{itm:increments} was repeated 4 times with $\left( t_\text{sim};T_\text{max} \right)$ having the values $\left(10^3;\, 5\cdot 10^1 \right)$, $\left(10^4;\, 5\cdot 10^2 \right)$, $\left(10^5;\, 5\cdot 10^3 \right)$ and $\left(10^6;\, 5\cdot 10^3 \right)$.
    When testing the sensitivity range, we compared the periods obtained for different values of $s^*$ (see step \ref{itm:test}) that were usually sampled in a roughly logarithmic fashion.
    In the rightmost column we provide a minimal and maximal value for $s^*$ that yield the same periods, as well as $f_{s^*}$, the largest factor between two successive values of $s^*$ within the sampled range in which period was consistent. In this step we used $T_\text{max} = 10^4$ except for the $307\times307$ square grid, where we used $10^3$.}
    \label{tab:simulation_parameters}
    \end{ruledtabular}
\end{table*}
We often chose the linear system size ($N$ in 1D and $L=\sqrt{N}$ in 2D) to be a prime number to avoid resonance of frequencies that divide the linear system scale.
We tested that the qualitative behavior of the system is not enforced by a specific odd or even side length (see Fig.~S1).
The effect of the order of magnitude of the linear system size can be seen in Fig.~\ref{fig:5_squ_pd}.

We generate initial conditions with average energy $\mu$ by assigning $z_i(0) = \mu \cdot r_i / \sum_i r_i$ for $i \in [1,2,...,N]$, where the real numbers $r_i$ are drawn from uncorrelated random distributions.
Unless otherwise specified we use a uniform distribution on the interval $[0,1)$, but we also test the robustness of our results by replicating some of the data with an exponential initial distribution: $p(r_i) = \exp(-r_i)$ for $r_i>0$ (see Fig.~S2).
Using this algorithm to generate initial conditions $\mu$ was sampled in equidistant steps $\delta \mu$ between $\mu_{\text{min}}$ and $\mu_{\text{max}}$ ({\it Details:} Tab.~\ref{tab:simulation_parameters}).

The simulations proceed using the following steps:
\begin{enumerate}
\item \label{itm:updates}compute a number $t_{\text{sim}}$ of sequential system updates starting with the initial state at $\mu_{\text{min}}$.
\item \label{itm:search}search for periodic attractors by computing up to $T_{\text{max}}$ additional updates and comparing each state generated at time $t\in[t_{sim}+1,t_{sim}+T_{\text{max}}]$ with the state at time $t_{\text{sim}}$.
If the locations of active sites, the "toppling pattern", of the states at $t_{sim}$ and $t$ match and the absolute difference in energy $|z_i(t) - z_i\left( t_{\text{sim}} \right) |<s$ $\forall i$ the state was classified as periodic.
$s$ hereby represents a sensitivity threshold (Tab.~\ref{tab:simulation_parameters}), which is explained later on.

\item \label{itm:orderpar}
If a periodic limit cycle was found, the period was set to $T = t - t_{\text{sim}}$ and one complete cycle was computed as a timeseries from $Z(t)$ to ${Z(t+T-1)}$, from which the order parameters $\langle a \rangle$ and $\langle \sigma \rangle$ were calculated.
As an estimate for the uncertainty, we computed the standard deviation of $a(t)$ and $\sigma(t)$ over said cycle are.

\item \label{itm:allmu} repeat steps \ref{itm:updates} and \ref{itm:orderpar} for all $\mu$ in the ensemble.

\item \label{itm:increments} repeat steps \ref{itm:updates} to \ref{itm:allmu} for all elements of the ensemble where no period was found with increased $t_{\text{sim}}$ and $T_{\text{max}}$ according to Tab.~\ref{tab:simulation_parameters}. If necessary, repeat these steps several times until a periodic attractor is found or computation time becomes too long. Always use the most recently calculated system update as initial state when continuing the simulation.

\item \label{itm:test} Test the robustness of the periods found under changes of the sensitivity $s$ by repeating step \ref{itm:search} with the respective final states for fixed $s$ and $T_{\text{max}}$.
Repeat this with different values of $s$.
If necessary, adjust $s$ to a convenient value $s^*$ and repeat steps \ref{itm:search} and \ref{itm:orderpar}.
We generally observe that the periods are independent of the exact value of $s$ in some range.
We provide a minimum range in Tab.~\ref{tab:simulation_parameters} under $s^*$.

\end{enumerate}
We generally observe that the Frobenius norm of the difference of states at $t$ and $t+T$
\begin{equation}
    ||\Delta_T(t)|| \equiv \sqrt{\sum_{i=1}^N \left( z_i(t+T) - z_i(t) \right)^2 }
\end{equation}
decays exponentially with $t$ (see Fig.~\ref{fig:2_transient}b), once the toppling pattern of the  limit cycle oscillation is reached, hence the periodic attractors are exponentially stable.
Justified by this observation, we assume that system states at times $t$ and $t+T$ become identical in the limit $t\to\infty$, if they are part of a period-$T$ cycle.
Therefore, we can safely stop the simulation once a period has been found.
Only in very rare cases we observed intermittent behavior, where the system seemed to converge towards a certain attractor over many iterations of its cycle of length $T$ until a deviation occurred on a single site that had been pushed towards the toppling threshold, which made the system "switch" the precedented attractor while still in transient.
The two attractors, however, only differed slightly in single locations of the toppling pattern and not significantly otherwise.
The design of step \ref{itm:increments} with a logarithmic increase of $t_\text{sim}$ and $T_\text{max}$ made the simulation significantly faster, because the transient times spanned several orders of magnitude with most transient times being small, but some systems not completing the transient within affordable computation times at all.
Similar things can be said for the periods that we found.
The rate of the above mentioned exponential decay towards the limit cycle, however, is finite and becomes smaller with increasing system size, which makes the introduction of the sensitivity $s$ necessary.
Therefore, however, the computed order parameters might not be exact but very close to the exact values of the limit cycle, with $s^*$ being a measure for how close.
The existence of finite lower and upper bounds on the sensitivity range can be explained as follows:
If $s$ is too small, i.e. close to machine precision or the exponential decay towards the attractor has not reached machine precision yet, numerical errors are made.
If $s$ is too large, different energy levels are counted as one, which can result in assigning an erroneous period.
Empirically motivated we set $s=4\times N \times \epsilon$, where $\epsilon$ is the machine precision of 64-bit floats, about $2.22\cdot 10^{-16}$.
This worked well in most cases, although for larger linear system scales ($N$ in 1D, $\sqrt{N}$ in 2D) larger values of $s^*$ were necessary due to slow convergence.

\section{\label{sec:results}Results}

\subsection{Characterizing the dynamics}
\noindent
 Our continuous sandpile model exhibits a great variety of dynamics despite the simple and deterministic update rule (Eq.~\ref{eq:toppling_rule}).
The transient behavior obviously depends on the detailed initial condition, 
but more interestingly, the model can also support many different periodic attractors --- even at a fixed average energy $\mu$.
To illustrate some of the complexity, let us begin by considering three simple periodic attractors on a finite 2D square lattice, all with an average energy $\mu=0.75$ (Fig.~\ref{fig:1_SimpleExample}).
In (a) all sites have energies below the critical threshold, $0\leq z_i < 1$ for all $i$, so there is no toppling activity.
We will refer to this as an absorbing state, a fixed-point, or a limit cycle (or attractor) with period $T=1$.
In (b) the sites are arranged into two two bipartite groups like the black and white squares on a checkerboard. In one group all sites have $z_i = 2\mu=1.5>1$ energy and in the other group all sites have zero.
The high energy sites are all above the toppling threshold, so after one timestep all the energy will be transferred to the other group.
After two timesteps, the energy is back where it started, so this alternating checkerboard pattern is a limit cycle with period $T=2$.
In (c) the sites are separated into diagonal lines with three distinct values: $0$, $\mu=0.75<1$ and $2\mu=1.5>1$. 
At every timestep, the high-energy sites topple and distribute all their energy to the neighboring sites.
The resulting state looks like the original state, but translated by one diagonal step.
This ``diagonal wave'' pattern is a limit cycle with period $T=3$ because the original state is recovered after three steps.
The three attractors all have constant activity $a(t)$ and (spatial) standard deviation $\sigma(t)$, but with very different values: $a(t)$ is zero, $1/2$, and $1/3$, respectively; and $\sigma(t)$ is 0.18, 1 and $\sqrt{2/3}\approx0.82$.

The three simple examples shown in Fig.~\ref{fig:1_SimpleExample} are a vanishingly small fraction of all the possible attractors of the continuous sandpile model.
Rather than enumerating possible attractors, the rest of this paper will, therefore, be focussing on describing the structure of `typical limit cycles'.
By `typical limit cycles', we mean the ones reached when the simulation is started with random initial conditions (see Methods).

\begin{figure*}
    \centering
    \includegraphics[width=.9\textwidth]{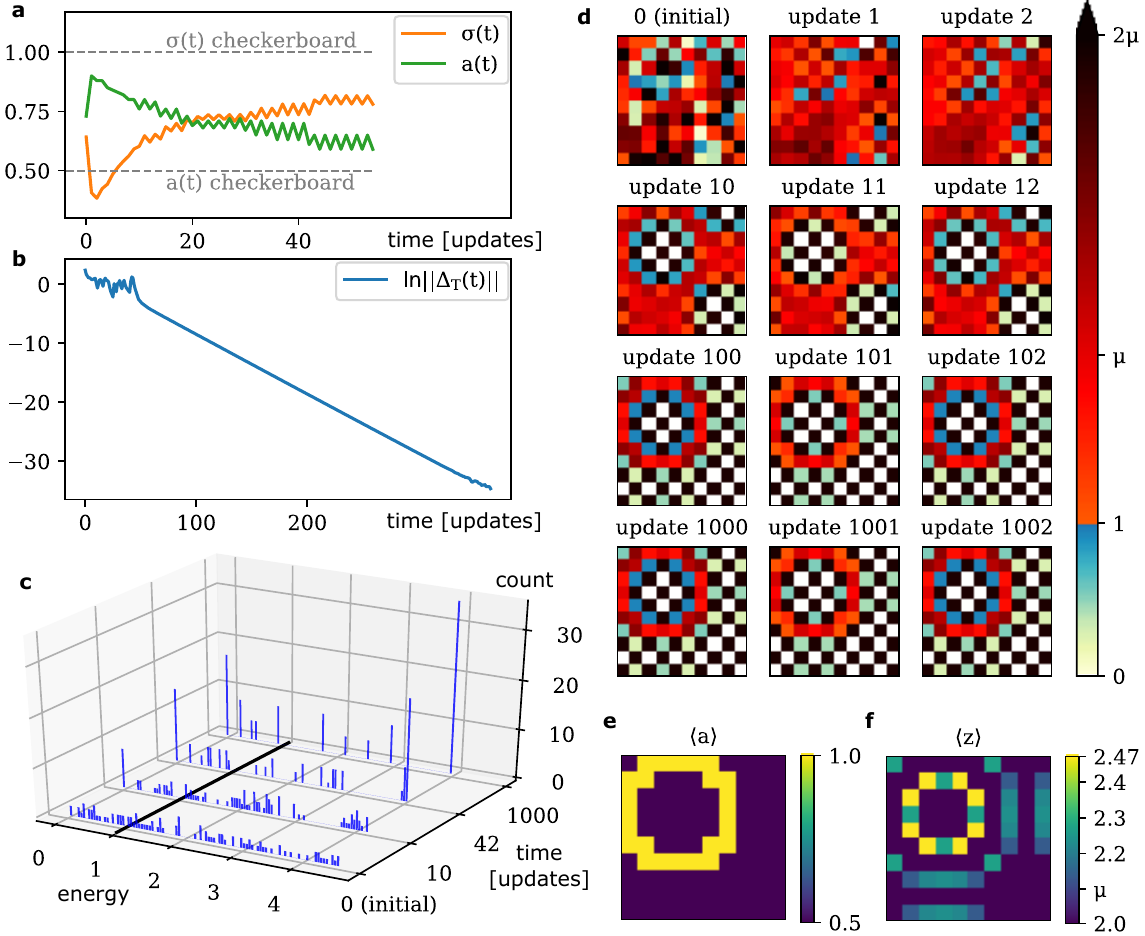}
    \caption{
    Transient behavior of an initial state with $\mu=2.1$ drawn from a uniform random distribution between 0 and $2\mu$ on a $10\times 10$ square lattice with periodic boundaries (torus).
    The toppling pattern of the limit cycle, which is a checkerboard with a circular boundary, or "frustrated furrow", and period $T=2$, is reached after 43 updates.
    (a) non-monotonic transient of the normalized standard deviation $\sigma(t)$ and activity density $a(t)$ that reaches a period-two oscillation after 43 updates (the asymptotic amplitude is reached after 45 updates). The upper (lower) gray line indicates $\sigma=1$ ($a=0.5$), which a perfect checkerboard without furrow would have;
    (b) exponential decay of $||\Delta_{T=2}(t)||$ after the toppling pattern of the limit cycle is reached;
    (c) Temporal evolution of the histogram of the site's energy content. The formation of sharp peaks begins before the toppling pattern of the limit cycle is reached and continues afterwards. The black line indicates the toppling threshold at unity;
    (d) snapshots of the system state $Z(t)$ during and after its transient show the emergence of a "checkerboard" with a circular furrow. The inner furrow sites (red squares in the panels at $t=100,\ldots1002$) are active in every update (see panel e) and have the same energy content as the checkerboard part (black and white sites) when averaged over one period (see panel f). The sites next to the furrow (green to blue or black) also topple in a spatio-temporal checkerboard pattern, but have a higher energy content than the checkerboard sites, since they are never completely empty (white).
    The ring-like structure that emerged during the transient interacts with itself across the periodic boundaries through these sites. This is not typically seen in larger systems like in Fig.~\ref{fig:6_squ_states};
    (e) spatial plots of the average activity in units of toppling events per update. The red furrows in panel d are active in every update, while all other sites topple every second update in a checkerboard-like pattern;
    (f) spatial plot of the average energy; The red furrow sites in panel d, updates 1000-1002, have the same energy content as the sites that belong to the checkerboard. }
    \label{fig:2_transient}
\end{figure*}

Let us consider an example of a typical trajectory with a random initial condition with a mean energy $\mu=2.1$ (Fig.~\ref{fig:2_transient}).
The initial activity $a(0)=(2\mu-1)/2\mu$ reflects the number of sites that are initially above the unit threshold.
Interestingly, activity then first increases further, which may at first be surprising, with the majority of sites toppling in the first timestep.
However, the probability of any of these sites to neighbor at least one other toppling site is $1-(1-a(0))^4$, hence $0.997$ in the example, for two or more toppling neighbors it is still $0.96$.
Therefore, many of the sites emptied at $t=0$ are immediately replenished to a level where they can topple again at $t=1$.
The result of such high initial activity is that the toppling rule in Eq.~\ref{eq:toppling_rule} is similar to a discrete diffusion operator, where each site always re-distributes its energy to the nearest neighbors --- an explanation consistent with the initial decrease in variance (Fig.~\ref{fig:2_transient}a). 
After the initial activity maximum, which is reached at $t=1$, activity again decreases, accompanied by an increase in variance.
Such increase in variance is documented by inspecting the histogram of site energies (Fig.~\ref{fig:2_transient}c), which becomes increasingly peaked at very low and high energies.
However, also several intermediate site energies occur more frequently than in the initial histogram ({\it compare} $t=0$ to $t=1000$).

Spatially, the initially random site configuration organizes into checkerboard-like $T=2$ patterns in several subregions (Fig.~\ref{fig:2_transient}d). 
These checkerboard regions are characterized by sites of $z_i=0$ surrounded by sites of large energies and these regions expand in space over time until most of the lattice is characterized by checkerboard-like patterning. 
Yet, contiguous subregions of energies exceeding the unit threshold remain.
For the system at hand it is found that the configuration at $t=1002$ is already very close to a numerically-perfect $T=2$ limit cycle (Fig.~\ref{fig:2_transient}), thus, we do not expect any further departure from this pattern in the $t\rightarrow\infty$ limit.
Re-inspecting the histograms (Fig.~\ref{fig:2_transient}c) it is worth noting that the maximum site energies reached in the periodic attractor are smaller than $2\mu = 4.2$.
This can be explained as follows:
If, a site's energy reached values of more than $2\mu$ in at least one state of a periodic attractor with $T=2$, its neighbors would have to compensate the high outflow of energy by achieving the same value at least once in a cycle, too. This argument can be repeated which leads to a contradiction between all sites having an average energy exceeding $\mu$ and $\mu$ being their average energy per definition.

Another emergent constraint exists: since (i) all sites must be active at some times and (ii) active sites must exceed the unit threshold, replenishment of any previously active site will always cause such a site to exceed the energy of $1/4$ --- resulting in an energy gap between zero and $1/4$ (more generally: $1/k$, with $k$ the coordination number of the lattice) in the emergent histogram (Fig.~\ref{fig:2_transient}c) and phase diagrams later on.
Another class of states, present on a sublattice, seems to exist: one where neighboring sites have similar energy which exceeds the unit threshold and simply exchange this energy at every timestep by repeated toppling (Fig.~\ref{fig:2_transient}d, red shades, and Fig.~\ref{fig:2_transient}e). 
Such high-activity, low-variance sub-lattices, which we here also refer to as "frustrated furrows", in analogy to the literature \cite{deutsch2005cellular}, appear to coexist with the $a=1/2$ checkerboard regions.
In the vicinity of the frustrated furrows average energy often appears to be greater (see Fig.~\ref{fig:2_transient}f), here allowing the system mean to equal $\mu=2.1$. 

\subsection{1D ring}
\begin{figure*}
    \centering
    \begin{overpic}{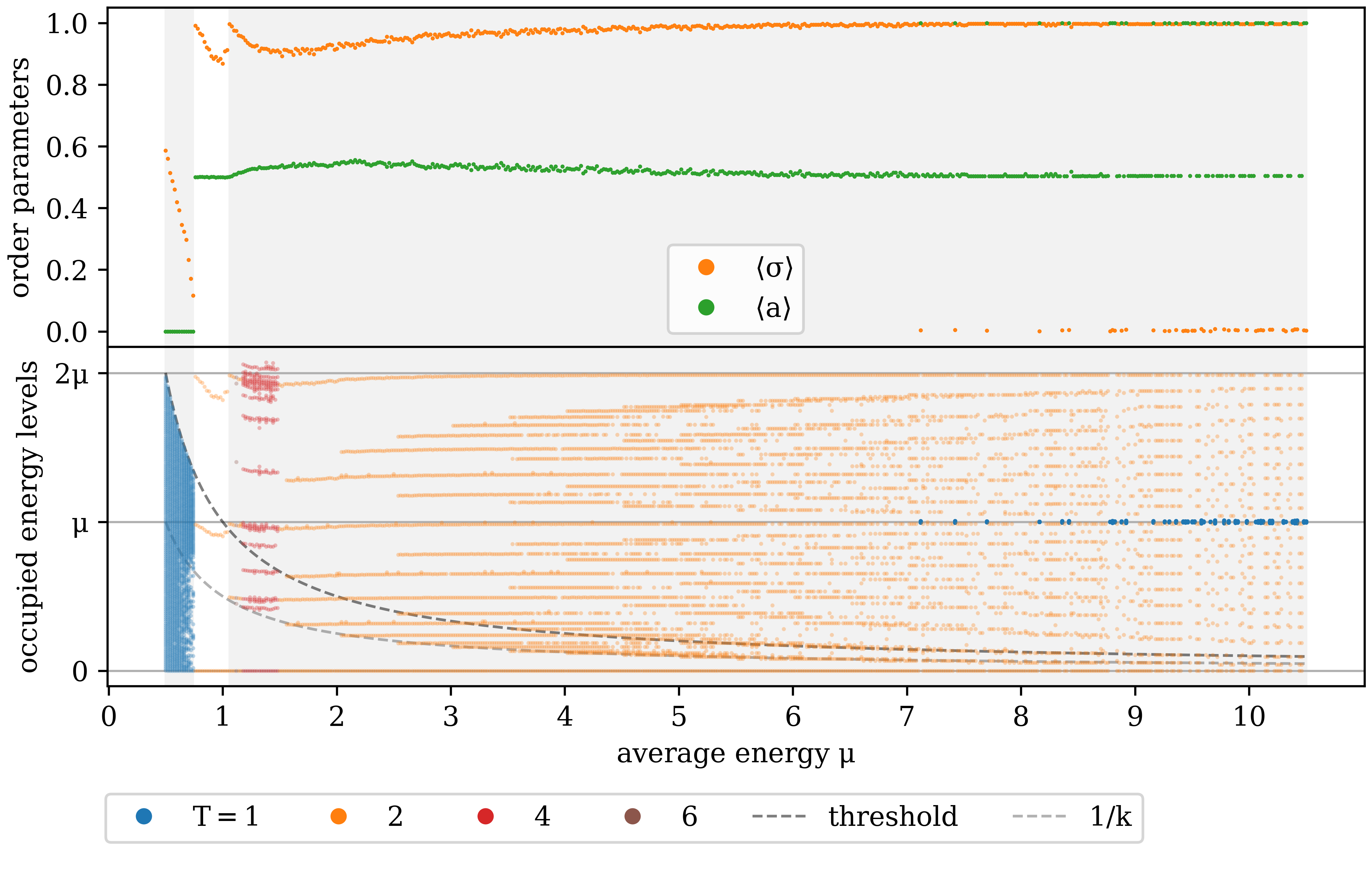}
       \put(12,65){\bf A}
       \put(15,65){\bf B}
       \put(20,65){\bf C}
    \end{overpic}
    \caption{Phase diagram displaying properties of the limit cycles found on a 1D linear chain (ring) of length $N=1997$ with coordination number $k=2$. Initial conditions were generated from a uniform random distribution and $\mu$ was sampled in steps of $\delta \mu = 0.02$ between 0.5 and 10.5. Further parameters can be found in Tab.~\ref{tab:simulation_parameters}.
    (top) different phases (shaded and white areas) can be distinguished by changes of slope or value of the order parameters $\langle \sigma\rangle$ and $\langle a \rangle$, which are the normalized standard deviation and activity density, averaged over a complete cycle. The respective standard deviations (error bars) are invisibly small. The phase transitions are located at 0.75 and 1.05 with an estimated uncertainty of $\delta \mu = 0.02$.
    (bottom) the occupied energy levels in a one state of the limit cycle (if found) for every sampled $\mu$ are binned into 150 evenly spaced bins and their occupancy is indicated by a data point.
    vertical gaps indicate that no period was detected within the simulation time and set parameters, hence this realization of the system is most likely still in transient.
    Using different states from the cycles did not alter the picture noticeably.
    Period is indicated by color. Each phase consists of one or several $\mu$-ranges that are governed by specific periods and occupied levels indicating limit cycles with large basins of attraction.
    Note that we detrend the diagram by scaling the energy-axis with $\mu$.}
    \label{fig:3_lin_pd}
\end{figure*}

\noindent
The example shown in Fig.~\ref{fig:2_transient} highlights the complexity of the spatial pattern arising for a single value of $\mu$.
Before discussing the corresponding phase diagram for the 2D square lattice, we turn to that of the less complex one-dimensional (periodic) ring lattice, where the coordination number is $k=2$ (Fig.~\ref{fig:3_lin_pd}).
Note that in this and all following phase diagrams we display one state from a periodic limit cycle that resulted from random initial conditions for each value $\mu$ that was sampled. This attractor state is likely to have typical characteristics and belong to a class of attractors with large basin of attraction size.
We observe that the phase diagram does not look significantly different if the simulation is repeated with a different random seed or if another state of the same attractor is displayed.
The order parameters displayed are averaged over one cycle of $T$ states and have invisibly small error bars.

For small average energies $\mu\lesssim0.8$ a phase (labeled A in Fig.~\ref{fig:3_lin_pd}) exists which becomes inactive within the limit $t\rightarrow \infty$.
The steady state is characterized by all site energies $z_i\leq 1$, $i\in \{1,\dots,N\}$, that is, $T=1$.
With increasing $\mu$ the fraction $a(0)=(2\mu-1)/(2\mu)$ of initially active sites must increase.
At $t=1$ the fraction of active sites $a(1)$ will also increase with $\mu$ because each of the $a(0)\cdot N$ sites will be able to activate one of the previously inactive $\left(1-a(0)\right)\cdot N$ sites.
In some cases, two active sites will be able to promote either a previously inactive site or a previously already active site to exceed the threshold.
Notably, as total activity $\sum_{t=0}^{\infty}a(t)$ increases with increasing $\mu$, fewer sites will remain completely inactive at all $t$ and the probability density of sites with energy $0<z_i<1$ will decrease. 

\begin{figure}
    \centering
    \includegraphics[width=.4\textwidth]{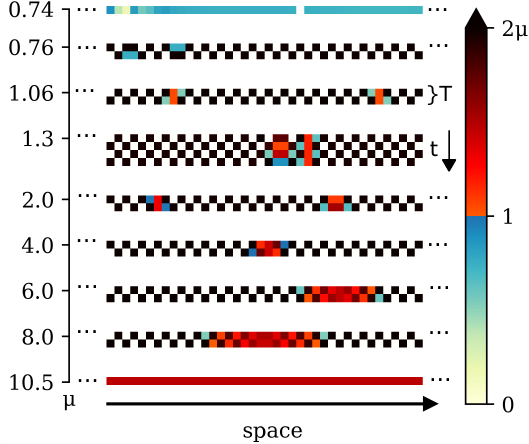}
    \caption{Periodic cycles for different $\mu$ on the 1D linear chain (ring) of length $N=1997$ with coordination number $k=2$ as in Fig.~\ref{fig:3_lin_pd}.
    Each block displays 40 sites of the chain.
    The dots ("\ldots") indicate that not all sites of the chain are shown.
    The corresponding $\mu$ is marked on the left axis.
    Subsequent rows show the same sites at the next update, which is indicated by the arrow on the right side of the fourth panel from the top.
    The number of rows in a block thus equals the period, which is indicated next to the third block from the top.
    Energy is indicated by color.
    The furrows that that separate the checkerboard pattern in all but the top and bottom block get larger with increasing $\mu$.
    Note that in the fourth block, where $T=4$ two different types of furrows are shown: The left one spans over four sites and is period-4, while the right one is identical to the period-two furrows in the block above that span 3 sites.}
    \label{fig:4_lin_states}
\end{figure}

When $\mu$ is increased further, we speculate that a type of percolation threshold is exceeded, beyond which activity affects all sites. 
Whereas we leave further thought of this threshold to future work, one path towards its exploration could be a directed percolation approach.
Even for the finite system, this transition appears discontinuous when quantified by the activity $\langle a\rangle$ or the spatial variance $\langle\sigma\rangle$.
For average energies in the parameter region B (marked in Fig.~\ref{fig:3_lin_pd}), the typical attractors are dominated by regions of clear checkerboard patterns separated by ``frustrated furrows'' as discussed above ({\it compare}: Fig.~\ref{fig:4_lin_states}).
These frustrated furrow states do not impact on overall activity, as they alternate between sequences $\{a,i,i,a\}$ and $\{i,a,a,i\}$, where $a$ and $i$ refer to active and inactive sites, respectively, thus leaving total activity constant.

At $\mu=1.05$, as the value of the middle energy level of phase B approaches and reaches the toppling threshold from below, a second phase transition occurs.
The parameter region C (marked in Fig.~\ref{fig:3_lin_pd}) is characterized by a multitude of much more complex states that involve more and more distinct energy levels with increasing $\mu$.
The overall dominance of the checkerboard pattern is not destroyed, but elaborate sequences of increasingly active sites emerge. 
Notably, the energy spectrum for a wide range of $\mu$ appears to be limited by the value $2\mu$ --- yet, exceptions exist for a range of $\mu$, where even this value is exceeded.
As explained earlier on, this is only possible for $T\neq2$
The population of additional energy levels with increasing $\mu$ can be explained qualitatively: 
even states at intermediate values $\approx \mu$ will now topple, since they exceed the unit threshold. Such toppling will create energy levels in the previously vacant region $[1/k,\mu]$.
When neighbors further topple into such sites, energy levels in the previously vacant region $[\mu,\approx 2\mu]$ are created, which explains the approximate symmetry of energy levels along the horizontal $z=\mu$ axis.
This dynamics continues to induce even more energy levels:
when $\mu$ is further increased, more and more of the lower energy levels will topple and give rise to further "satellite" energy levels --- eventually opening up for a near continuum of states.
These levels eventually become so finely-spaced, that it is hard to resolve them in the numerical simulations.
For very large $\mu\approx 10$, we observe $\langle a\rangle =1$ and $\langle\sigma\rangle=0$, that is, the operator in Eq.~\ref{eq:DFES_toppling_rule} acts as a diffusion operator, which removes all variance at sufficiently large $t$.

Most phase transitions take place within in a $\mu$-range that is small compared to the range between the phase transitions at the lowest and highest $\mu$.
An exception from this rule of thumb is the "smeared out" transition at the highest $\mu$ to the phase characterized by $\langle a \rangle = 1$ and $\langle \sigma \rangle = 0$. 
For the sake of clarity, we only show the order parameters and occupied energy levels of the attractor state reached from a single random initial state for each sampled value of $\mu$.
Both $\langle a \rangle$ and $\langle \sigma \rangle$ do not jump back and forth.
Instead, they can reach both values depending on the exact initial state at $t=0$.
The above-mentioned phase transition happens within a small $\mu$-range, however, if one of the following conditions is met: 
(i) the underlying network contains triangles, like the triangular lattice and linear chain with next-to-next neighbor interactions in Fig.~S4; 
(ii) if strong noise is present, as in Fig.~S7 (bottom).
To sum up, the observed phase transitions are of stochastic nature due to the strong dependence on initial conditions, but in most of $\mu$-space there is only one dominating attractor state for the ensemble of initial conditions that we use.

Fig.~\ref{fig:4_lin_states} shows parts of the periodic limit cycles found in the simulation for certain values of $\mu$.
One can see that with increasing $\mu$ the furrows get broader and consist of sites with more and more different energy levels, which corresponds to the creation of new energy levels with increasing $\mu$ as discussed above.
A heuristic reason for the broadening is the higher fraction of sites that topple during the first update, which makes them more likely to continue toppling as opposed to fewer toppling sites that are scattered more scarcely and are more likely to end up in a checkerboard pattern.
The simplest non-trivial type of attractor belongs to phase B in Fig.~\ref{fig:3_lin_pd} and is shown in the second panel in Fig.~\ref{fig:4_lin_states} from the top, that is, at $\mu =0.76$.
It is equivalent to a rescaled version of the DFES model, which is defined with an integer-valued field and the BTW toppling rule of "giving one grain to each neighbor" when toppling. The toppling threshold for these models is usually the coordination number $k$.
The equivalence becomes clear when rescaling the energy levels that are roughly at 0, $\mu$ and $2\mu$ to 0, 1 and 2.
An analytical solution as in \cite{Dall_Asta_2006} is now possible.
Given a period of $T=2$, the constraints on the structure can be derived:
Every site has to topple --- and thus have an energy of 2 --- exactly once in a cycle, that is, every second update, because otherwise, if one site would topple 0 (2) times, its neighbors and hence every other site would have to topple 0 (2) times in a cycle.
Furthermore, sites with energy 1 can only appear within the sequence {2,1,1,2}, sites with energy 0 only within the sequence {2,0,2} and sequences like 2,2,2 or longer are forbidden.
Note that this dynamics only works if $\mu\leq1$ so that only the top level topples.
As expected, the corresponding phase B terminates at $\mu=1$.
In contrast to \cite{Dall_Asta_2006} the period $T$ does not have to divide $N$ here, because the highest energy level that is usually allowed in BTW DFES, which is $2k-1=3$, is not occupied in our rescaled version.

Simulations with several different random initial conditions drawn from the same distribution at each $\mu$ also imply that the dominance of certain limit cycles and the phase transitions for given $\mu$ are of stochastic nature because of strong dependence on initial conditions.
A checkerboard, for example, is a stable periodic state for $\mu \in (0.5,\, \infty)$, but is not the dominant attractor in the entire range of average energies.

\subsection{2D square lattice}
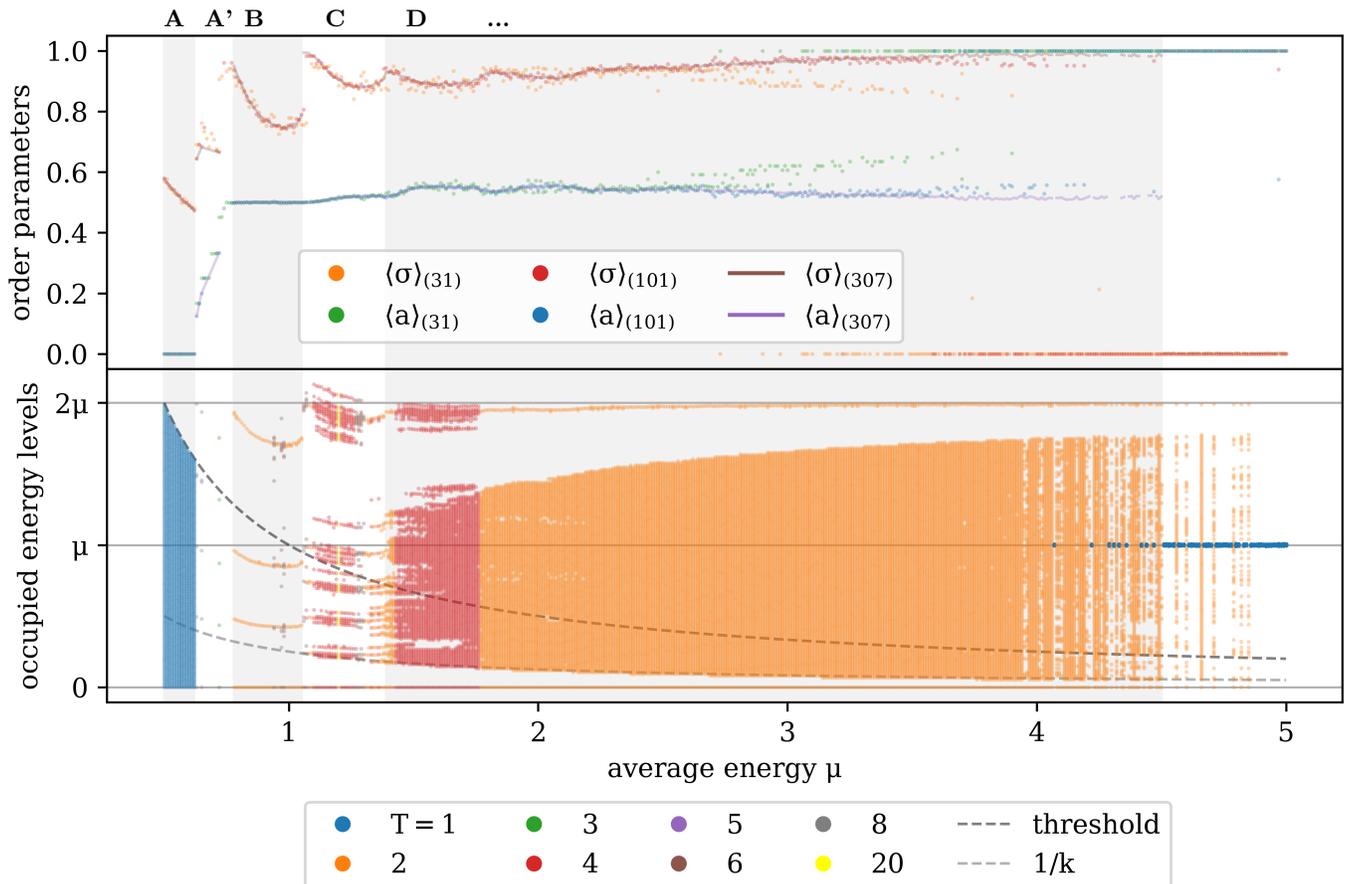
\begin{figure*}
    \centering 
    \begin{overpic}{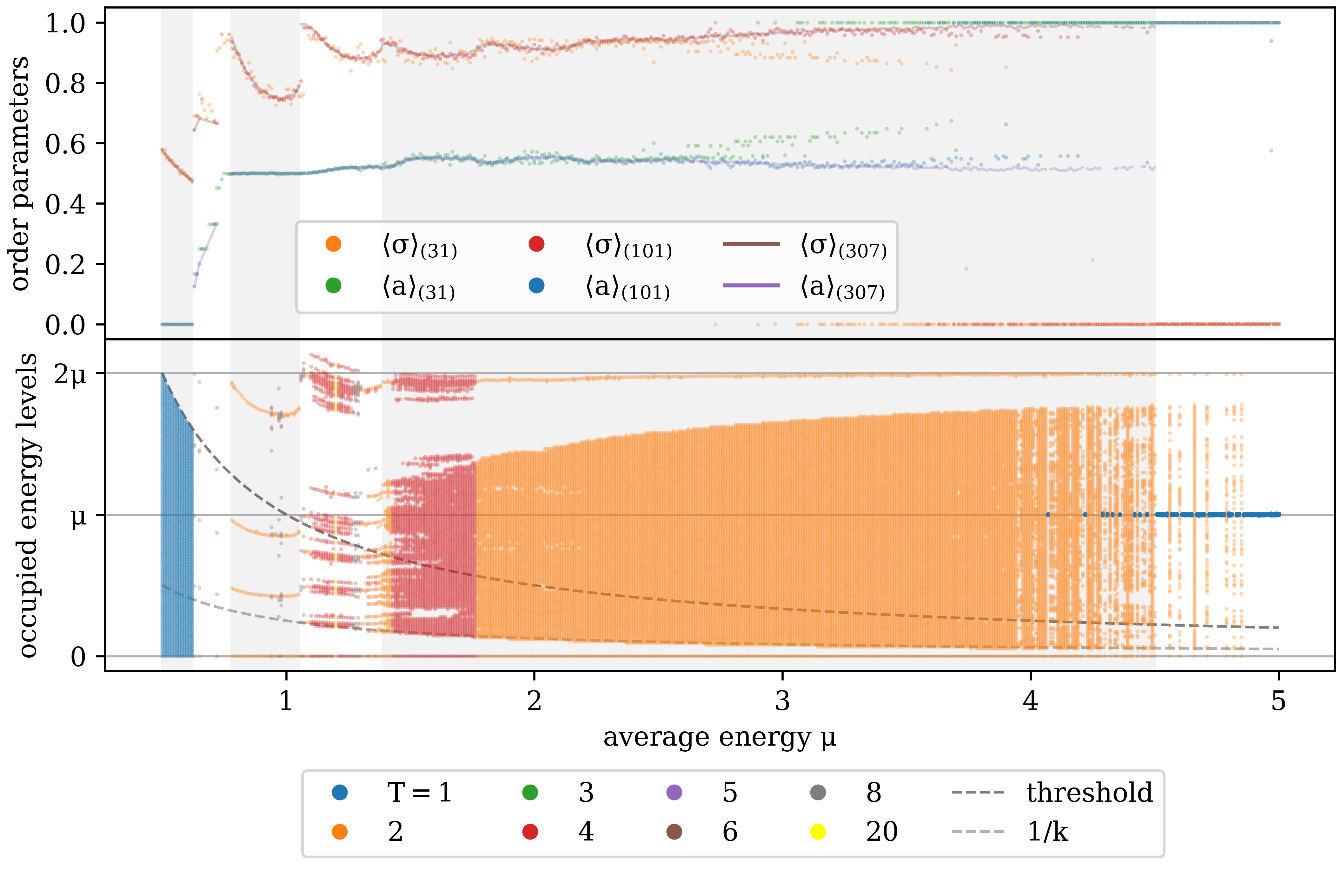}
       \put(12,67){\bf A}
       \put(15,67){\bf A'}
       \put(18,67){\bf B}
       \put(24,67){\bf C}
       \put(30,67){\bf D}
       \put(36,67){\bf ...}
    \end{overpic}
    \caption{Phase diagram displaying properties of the limit cycles found on a 2D square lattice (torus) of side lengths $\sqrt{N}=31$, 101 and 307 with coordination number $k=4$. Initial conditions were generated from a uniform random distribution and $\mu$ was sampled in steps of $\delta \mu = 0.01$ between 0.5 and 5.0. Further parameters can be found in Tab.~\ref{tab:simulation_parameters}.
    (top) $\langle \sigma\rangle$ and $\langle a \rangle$ are shown for different $\sqrt{N}$. The respective standard deviations (error bars) are invisibly small.
    Different phases are indicated by shaded and white areas.
    The curves become smoother with increasing system size.
    The phase transitions (determined for the $307\times 307$ system) are located at $\mu\in \left\{0.625, 0.775, 1.055, 1.385, 4.505\right\}$ with an estimated uncertainty of $\delta\mu = 0.01$.
    (bottom) the occupied energy levels in a randomly selected state of the limit cycle (if found) on the $N=307\times 307$ torus for every sampled $\mu$ are binned into 150 evenly spaced bins and their occupancy is indicated by a data point. Period is indicated by color. Each phase consists of one or several $\mu$-ranges that are governed by specific periods and occupied levels indicating limit cycles with large basins of attraction.}
    \label{fig:5_squ_pd}
\end{figure*}

\begin{figure*}
    \centering
    \includegraphics{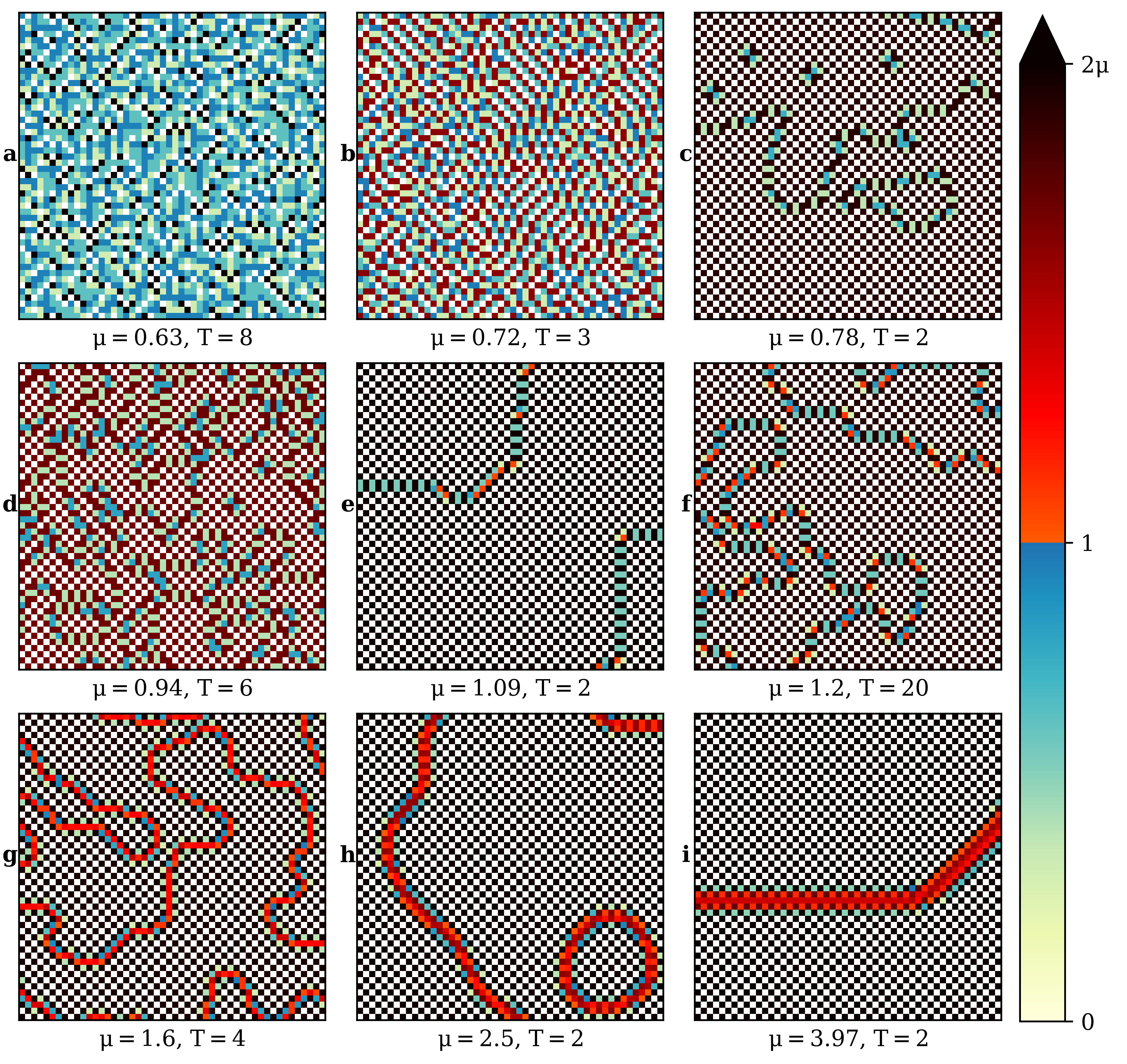}
    \caption{Snapshots of parts of a randomly selected state of the limit cycle for different $\mu$ on the $N=307\times 307$ 2D square lattice from Fig.~\ref{fig:5_squ_pd}.}
    \label{fig:6_squ_states}
\end{figure*}

\noindent
We now turn to the phase diagram for the 2D square lattice (Fig.~\ref{fig:5_squ_pd}).
For low mean energy (phase A), $\mu\lesssim0.5$, inactive states are again present, with the toppling of several sites leading to a reduction in initial variance $\langle \sigma\rangle$. 
Yet, whereas $\langle \sigma\rangle$ appeared to approach zero for the 1D ring (Fig.~\ref{fig:3_lin_pd}) in the case of the 2D square lattice it now only decreases until reaching a finite positive value (Fig.~\ref{fig:5_squ_pd}). 
When $\mu$ is increased further, a new phase A' is entered, of which there is no analog on the 1D ring: whereas the activity increases in a step-like manner, the spatial variance is increased compared to the initial distribution and to phase A, but jumps erratically as a function of $\mu$.
This phase is characterized by $T$ decreasing in unit steps from 8 to 2 \footnote{Note that $T=7$ has not been found in the simulations (see Fig. \ref{fig:5_squ_pd}).
More extensive simulations, perhaps also using larger systems, could reveal if the cascade can start at $T>8$ and include $T=7$, which is, however, not the scope of this work.} while the activity density increases as $\langle a \rangle = 1/T$, which is reminiscent of the "devil's staircase" found in \cite{jensen1983complete}.
Fig.~S6 provides a closer look at this phase and gives an impression of the wide spectrum of possible limit cycle periods within the a small $\mu$ range thereafter.
The corresponding states to phase A' that form periodic cycles of length $T$ consist of 5 evenly spaced energy levels that span from 0 to $\approx 1.26$.
Every site topples exactly once during a cycle.
Spatially, a complex pattern results (Fig.~\ref{fig:6_squ_states}, $\mu=0.72$), where subregions show patchy structure that is often periodic at small length scales.
Yet, the overall pattern can be characterized as strongly frustrated, with coexistence of many competing phases.
When $\mu$ is further increased to $\mu\approx0.78$ a dominant checkerboard pattern emerges, $\langle a \rangle\approx0.5$ and $\langle \sigma \rangle\approx 1$, which is interspersed by "furrow"-like patterns showing alternations between super and sub-threshold sites.
As $\mu$ is increased, the area covered by furrows increases, reducing $\langle \sigma\rangle$.
Further, the contrast in the remaining checkerboard is reduced, with active sites now reaching values far below $2\mu$.

This contrast is abruptly restored in the subsequent phase, where the fraction of space covered by furrows is again reduced.
In addition, the furrows now contain regions of continuous activity, which manifests itself in a slight increase of $\langle a \rangle$ beyond the value of $1/2$.

Further phases can be distinguished as $\mu$ is again increased. 
Generally, furrows tend to widen, as was the case in 1D, and appear to decrease in curvature, leading to more and more elongated furrow structures. 
The activity within furrows tends to be larger than $1/2$.
At large $\mu$ the system again reduces to a homogeneous, fully-active state, where no more structure is present.

As was the case for phase B on the 1D ring, phases A' and B on the 2D square lattice are equivalent to a rescaled DFES with the BTW toppling rule and the additional constraint that only the $k+1$ levels from 0 to $k$ are occupied.
As in \cite{Dall_Asta_2006} each site topples once every $T$ updates, but as in 1D $T$ doesn't have to divide the lattice's side length $L$.

\subsection{Sensitivity to domain size, initial distribution and geometry}
\noindent
Fig.~\ref{fig:5_squ_pd} (bottom) only shows the occupied energy levels of the largest system size, $N=307\times307$.
The phase diagrams for different system sizes do not differ significantly, but finite size effects become less noticeable with increasing system size.
To assess whether the details of the phase diagram might be induced by lattices sizes commensurate with the phases found, we compare square lattices of $L=100$ vs. $L=101$ linear dimension (Fig.~S1), finding that the differences are insignificant.
In particular, the general features of the phase diagram are preserved, including the "devil's staircase"-like feature discussed above.

In Fig.~S2 we explore the sensitivity to the choice of initial probability distribution: using an exponentially-decaying probability density function with the same mean as the previous uniform distribution to draw initial states, we again obtain overall similar results.

When changing the lattice geometry to be a honeycomb ($k=3$), a somewhat altered phase diagram results (see Fig.~S3). Yet, the devil's staircase-like feature and the appearance of discretely spaced vs. more continuous energy levels are still visible. 
This also goes for the triangular lattice, where a rather large parameter region of period-three states emerges, which is characterized by $\langle a\rangle = 1/3$.
The resulting phase diagram is similar to the one of a 1D ring with next and next to next neighbor interactions, hence $k=4$, despite their different dimension (see Fig.~S4).
Both lattices contain triangles and thus do not feature bipartiteness.
In contrast to boundary-induced frustration, which still resulted in the formation of checkerboard-like patches separated by furrows on the square grid, this local prohibition of bipartiteness makes checkerboard patterns impossible.
Therefore, the intermediate $\mu$ range is dominated by attractors with fewer energy levels, while the low and high $\mu$ regimes resemble the phase diagrams of bipartite lattices.
Examples of the corresponding states, which are more complex than checkerboards with furrows, are shown in Fig.~S5.
A remarkable difference to bipartite lattices is that the continuous phase takes over gradually by covering more and more of the lattice sites.

\subsection{Robustness against noise}
\begin{figure*}
    \centering
    \includegraphics{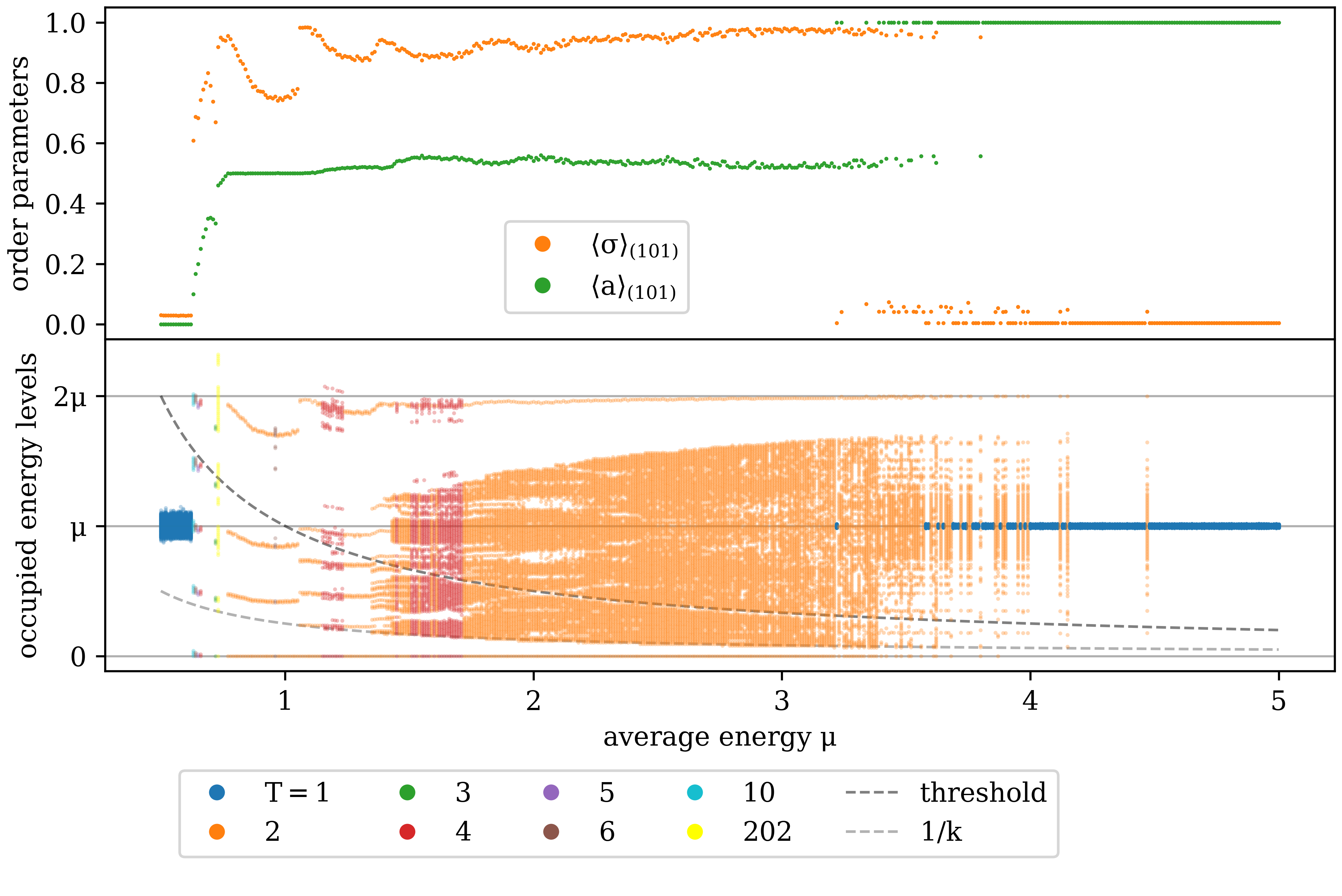}
    \caption{Phase diagram for the model on a $101\times 101$ square lattice with periodic boundaries showing the effect of noise intensity $\varepsilon = 0.01$.
    There are no significant differences to Fig.~\ref{fig:5_squ_pd}.
    Compare also to Fig.~S1 (bottom).
    The most striking difference is that with noise the inactive absorbing attractor state in the low $\mu$ regime is nearly homogeneous, hence $\sigma \approx 0$.
    Period is here detected by looking for a toppling pattern that repeats itself 10 times.}
    \label{fig:7_squ_101_noise}
\end{figure*}

\noindent
In this section we show that applying weak to intermediate noise in every update does not alter the behavior of the system significantly, which implies that the obtained results are not just induced by a fine-tuned choice of geometry, the toppling rule and boundary conditions (see Fig.~\ref{fig:7_squ_101_noise}).
The noise was implemented as follows: before each update, every site $i$ distributes a total of $r_i \cdot \varepsilon \cdot z_i$ of its energy $z_i$ at equal shares to its neighbors, where $r_i$ is a random number between 0 and 1 that is drawn independently for each site in each timestep and the real number $\varepsilon\in[0,1]$ is the noise intensity.
In other words, between each two deterministic updates according to Eq.~\ref{eq:toppling_rule} we insert an update according to a modified toppling rule in which the threshold-inducing $\Theta$ function from Eq.~\ref{eq:toppling_rule} is replaced with uncorrelated noise, that is,
\begin{equation}
    \label{eq:noise}
    z_i(t) \to z_i(t) \cdot \left[1 - \varepsilon \cdot r_i \right]
    + \sum_{j\in\mathcal{N}_i} \frac{z_j(t)}{k_j}\cdot \varepsilon \cdot r_j.
\end{equation}
Motivated by our previous empirical findings we chose $t_\text{sim}=233,\!100$, after which almost all trajectories had converged towards a periodic orbit in the simulations on the same lattice without noise.
This is twice the cumulative simulation time from those simulations, if step \ref{itm:increments} is repeated three times with $t_\text{sim} \in \left\{10^3,\,10^4,\,10^5 \right\}$ and $T_\text{max} \in 5 \cdot\left\{10^1,\,10^2,\,10^3 \right\}$ respectively.
Periods, which are marked by color in the bottom panel of Fig.~\ref{fig:7_squ_101_noise}, are here detected using a different method:
after the 233,100 updates, that we assumed to be sufficient for the transient, we computed a timeseries by simulating $10^4$ additional updates and compared toppling patterns.
If the sequence of the $T$ last states in the timeseries yielded a toppling pattern that repeated itself 10 times, therefore $T_\text{max}=1000$, at the end of the timeseries, we assigned the timeseries the period $T$ --- $T$ being the smallest positive integer with that property.
Out of 451 sampled values of $\mu$ the period of the detected attractors differs in 19 cases when comparing the simulations on the $101\times101$ square lattice with the same initial states but with and without noise.
We also tested the effect of weaker and stronger noise (Fig.~S7).
Under $\varepsilon=0.001$ the phase diagram is even closer to the one without noise ($\varepsilon=0$).
Under $\varepsilon=0.1$ the emerging structure is altered noticeably but still conserved in some aspects.
We expect it to eventually break down under further increase of the noise $\varepsilon$.
It is remarkable that the energy levels remain sharply defined even under the presence of substantial noise.
We conclude that the conditional toppling operator $\hat{T}$ in Eq.~\ref{eq:updating_operator} counteracts noise, which in turn acts as an unconditional diffusion operator. 

\section{\label{sec:discussion}Discussion and Conclusion}
\noindent
Many real-world phenomena are characterized by fluxes of continuous, rather than discretized, observables, quantities we generically refer to as "energy" throughout this study. 
Our model hence allows for smooth probability distributions of energy. However, the energy values that actually occur within the exponentially stable (see Fig.~\ref{fig:2_transient}) attractors of the dynamics, are highly constrained --- or "quantized."
These values are a result of the dynamical process defined by the model and are strongly dependent on the mean energy $\mu$. 
To be precise, they are a result of a complex reorganization process, by which the mean energy is made compatible with the overall dominant checkerboard-like spatio-temporal attractor of period two, if the lattice is free of triangles.
Impurities, constituted by more complicated soliton-like multi-site "particles" emerge, to compensate for the lack of commensurability.
To describe the spectrum of energy values as well as the dynamics, as a function of the tuning parameter $\mu$, for each lattice geometry we computed the phase diagram, that is, the positions of occupied energy levels, the activity and the spatial variance.
Even for the 1D ring an elaborate phase diagram results (see Fig.~\ref{fig:3_lin_pd}), which, due to the self-consistency between spatial organization and energy-level discretization, challenges analytical approaches.

The dynamics in Eq.~\ref{eq:toppling_rule} have been studied by Zhang with dissipative boundaries and slow insertion of energy into the system \cite{Zhang}.
Therein, a self-organized critical state arises with critical energy density, i.e., the point between dying out and ongoing activity, has been found at $\mu_c = 0.62 \pm 0.01$ for the 2D square lattice.
Analogous to the discrete BTW sandpile \cite{BTW}, the self-organized state of the 1D version is homogeneous with $\mu_c = 1$.
In 1D, our energy-conserving model is not as trivial and we find the absorbing-state phase transition between $\mu=0.74$ and 0.76.
In 2D, our findings coincide with those of Zhang, where we find that $\mu_c$ lies between 0.62 and 0.63.
This is also in agreement with an analytical value, $\mu_c \approx 0.6204$, obtained using a self-consistency approach for the original Zhang model with dissipative boundaries \cite{PTZ91}.
For the DFES on a 1D hypercubic lattice ($\mathbb{Z}$) the self-organized critical energy density in the dissipative case equals the density at the absorbing-state phase transition in the fixed-energy case \cite{meester2005connections}.
It has been proved numerically and with rigorous arguments, however, that this \textit{density conjecture} is not generally true (see \cite{Fey_2010} for the square lattice $\mathbb{Z}^2$ and other graphs), although the deviations are sometimes small.
Grassberger and Manna pointed out that the density at the absorbing-state phase transition depends on the initial conditions \cite{grassberger:jpa-00212432}.
Therefore, we do not expect that the densities match up exactly in the continuous model in 2D.

In the 2D Zhang model, the energy landscape of the self-organized critical state is on average homogeneous and isotropic in space and has four equidistant sub-threshold peaks that lie at zero and multiples of some energy quantum $E_0$ \cite{Zhang}.
Except for the peak at zero they have a finite spread, because of dynamical fluctuations and finite system size.
In the sub-critical case, $\mu \leq \mu_c$, J\'anosi found $2d - 1$ peaks (the peak at zero is missing there) and suggested a spacing of $E_0 = \mu_c/d$ for hypercubic lattices in $d$ dimensions \cite{JaIm90}.
L\"ubeck argued that on these lattices, which have coordination number $k=2d$, there are $k$ such peaks in the critical state and the energy quantum is given by $E_0 = \frac{k+1}{k^2}$ \cite{simZhang}.
That the number of peaks is given by the coordination number has also been suggested by D\'{\i}az-Guilera \cite{Guilera92} after simulations on different lattices.
If the dynamics in our fixed-energy version is simulated for sufficiently many timesteps, the energy landscape that emerges from random initial conditions is sharply peaked, but not homogeneous and isotropic in space due to the frustrated furrows as can be seen from Figs.~\ref{fig:2_transient}d and f, \ref{fig:4_lin_states}, \ref{fig:6_squ_states} and S5.
We speculate that it is, if one averages over an ensemble of initial conditions with fixed $\mu$ that is spatially homogeneous and isotropic itself.

The first phase after the absorbing-state phase transition (B in Fig.~\ref{fig:3_lin_pd}, A' in Fig.~\ref{fig:5_squ_pd}) seems to resemble the self-organized critical state with the exception that it has a single peak above the toppling threshold and thus $k+1$ peaks in total. 
We find that an argument similar to the one by L\"ubeck can be applied and yields the same expression for $E_0(k)$.
Combining the previous two expressions for $E_0$ we find $\mu_c = \frac{k+1}{2k}$.
This is in excellent agreement with the locations of the absorbing-state phase transitions that we find on the 1D linear chain ($k=2$), the honeycomb lattice ($k=3$, see Fig.~S3) and the 2D square grid ($k=4$).
On the triangular lattice and the linear chain with next-to-next-neighbor interactions (Fig.~S4) the formula underestimates the observed values by about 6\% and 10\% respectively.
The high-$\mu$ end of phase A in our phase diagrams resembles the broad peaks and the absence of the energy peak at $z=0$ in the sub-critical regime in \cite{JaIm90}. 
This can be seen particularly well in Figs.~S4 and S6.

When compared to the BTW DFES, which we refer to as the discrete model, our continuous model has the same stabilisation criterion \cite{FdBMR09, FLP10}.
If sites are updated successively, our model does not have the Abelian property in the inactive phase.
Inspite of the higher complexity and the much larger phase space that come with a real-valued field,
the dynamics still drive the system state towards limit cycle oscillations with short periods as is the case for the discrete model \cite{Bagnoli_2003}. This, together with the dependence on initial conditions, shows the strong non-ergodicity of the dynamics.
In some regions of phase space the dominating attractors are in structure and dynamics equivalent to rescaled attractors of BTW DFES, for which an exact solution has been found in 1D \cite{Dall_Asta_2006}.
Accordingly, in these regions the phase diagram of our model resembles key features of the BTW DFES like the absorbing-state phase transition and "devil's staircase" \cite{Bagnoli_2003}.
The continuous model, however, shows flexibility against boundary-induced frustration effects as the behavior on square lattices with a side length of 100 and 101 does not show significant differences.
In fact, with increasing system size ideal checkerboard states become untypical even on bipartite lattices, where checkerboard-like patches with "frustrated furrows" as boundaries are preferred.
These furrows resemble results of activation-inhibition dynamics \cite{artavanis1999notch, deutsch2005cellular}, however, the presence of noise is not necessary for their formation and the corresponding attractor states are typically dynamical and oscillate between several states of this structure.
The furrows show a larger activity but the same average energy content as the checkerboard patches and sites in their vicinity tend to accumulate average energy above $\mu$.
When lattices with triangles are introduced, checkerboard patterns are geometrically frustrated. 
The checkerboard is then necessarily replaced by more complicated structures and the transition to homogeneous active absorbing states in the high $\mu$ regime becomes continuous (see Fig.~S4).

Furthermore, with increasing $\mu$ we find a multiplicity of phase transitions and an "explosion" of occupied energy levels that come with a broadening and straightening of the furrows in the attractor states.
Our results are within a reasonable range robust against the choice of system size, initial energy distribution and against noise (see Fig.~\ref{fig:7_squ_101_noise}). 
This shows that the observed behavior is not purely geometric or tied to purely deterministic updating.




\begin{acknowledgments}
\noindent
The authors gratefully acknowledge funding from the Villum Foundation (grant no. 13168). 
J. O. H. acknowledges funding from the European Research Council under the
European Union's Horizon 2020 Research and Innovation programme (grant no. 771859) and the Novo Nordisk Foundation Interdisciplinary Synergy Program (grant no. NNF19OC0057374).
J. N. thanks the Erasmus+ program.
\end{acknowledgments}

\bibliography{apssamp}

\clearpage
\includepdf[pages=1]{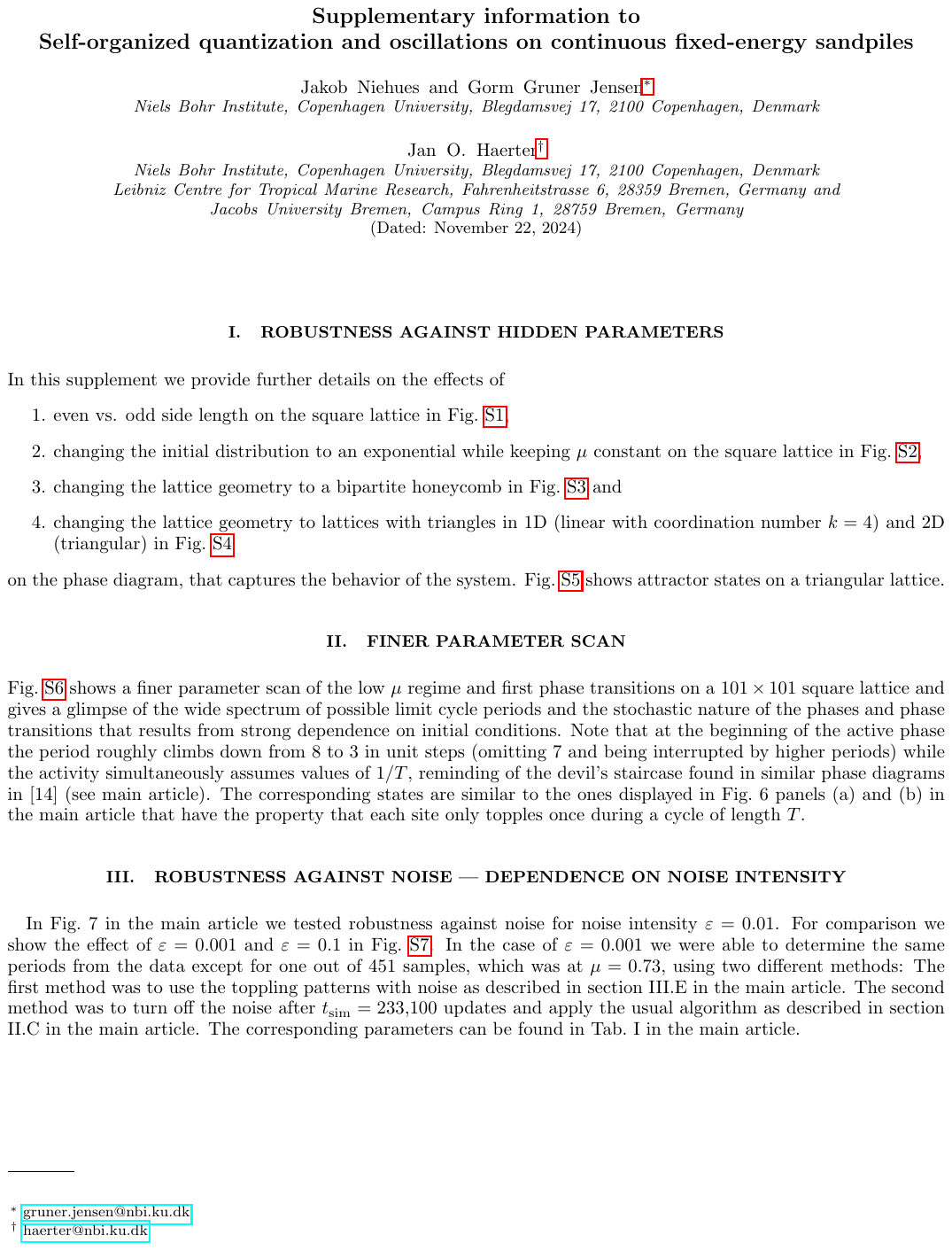}
\clearpage
\includepdf[pages=2]{supplement_arxiv_update.pdf}
\clearpage
\includepdf[pages=3]{supplement_arxiv_update.pdf}
\clearpage
\includepdf[pages=4]{supplement_arxiv_update.pdf}
\clearpage
\includepdf[pages=5]{supplement_arxiv_update.pdf}
\clearpage
\includepdf[pages=6]{supplement_arxiv_update.pdf}
\clearpage
\includepdf[pages=7]{supplement_arxiv_update.pdf}
\clearpage
\includepdf[pages=8]{supplement_arxiv_update.pdf}

\end{document}